\documentclass[twocolumn,superscriptaddress,prd]{revtex4}
\usepackage{graphicx}
\usepackage{dcolumn}
\usepackage{bm}
\usepackage{amsmath}
\usepackage{amsfonts}
\begin{document}

\title{Spherically symmetric static spacetimes in vacuum $f(T)$ gravity}
\author{Rafael Ferraro}
\email{ferraro@iafe.uba.ar}
\thanks{Member of Carrera del Investigador Cient\'{\i}fico (CONICET,
Argentina)} \affiliation{Instituto de Astronom\'\i a y F\'\i sica
del Espacio, Casilla de Correo 67, Sucursal 28, 1428 Buenos Aires,
Argentina} \affiliation{Departamento de F\'\i sica, Facultad de
Ciencias Exactas y Naturales, Universidad de Buenos Aires, Ciudad
Universitaria, Pabell\'on I, 1428 Buenos Aires, Argentina}
\author{Franco Fiorini}
\email{franco@iafe.uba.ar} \affiliation{Instituto de Astronom\'\i
a y F\'\i sica del Espacio, Casilla de Correo 67, Sucursal 28,
1428 Buenos Aires, Argentina}

\begin{abstract}

We show that Schwarzschild geometry remains as a vacuum solution
for those four-dimensional $f(T)$ gravitational theories behaving
as ultraviolet deformations of general relativity. In the gentler
context of three-dimensional gravity, we also find that the
infrared-deformed $f(T)$ gravities, like the ones used to describe
the late cosmic speed up of the Universe, have as the circularly
symmetric vacuum solution a Deser-de Sitter or a BTZ-like
spacetime with an effective cosmological constant depending on the
infrared scale present in the function $f(T)$.

\end{abstract}

 \maketitle



\section{Introduction}

$f(T)$ theories \cite{Nos,Nos2,Nos3,fdet1,Linder} have been
proposed in the last few years as an alternative to $f(R)$
modified gravity \cite{Sotiriou}, and they constitute a very
promising research area as is witnessed by the increasing interest
in the field
\cite{fdet2,fdet3,fdet4,fdet5,fdet51,fdet52,fdet53,fdet6,fdet7,fdet8,
fdet9,fdet10,fdet11,fdet12,fdet13,Miao2,fdetB1,fdetB2,fdet14,fdet15}.
The dynamical object in $f(T)$ theories is the field of frames
(vierbeins or tetrads), which involves the basis $\{e_a(x)\}$ of
the tangent space and its respective co-basis $\{e^a(x)\}$:
\begin{equation}
e^a_\mu\,e_b^\mu\ =\ \delta^a_b\, ,\ \ \ \ \ \ \ \ e^a_\mu\,e_a^\nu\
=\ \delta^\nu_\mu\, .\label{dual}
\end{equation}
The Lagrangian density
\begin{equation}
{\cal L}\,\propto\, e\, f(T)
\end{equation}
is built with the product of the determinant $e$ of the matrix
$e^a_\mu$ (the components of the forms $e^a$ in a coordinate basis)
times a function of the scalar $T$, which is quadratic in the set of
2-forms $T^a\, \equiv\, de^a$. The components $T^c(e_a,e_b)=T^c_{\
\mu\nu}\, e_a^\mu\, e_b^\nu$ are directly related with the
commutator of the $e_a$'s,
\begin{equation}
[e_a\, ,\, e_b]\ =\ T^c(e_a,e_b)\ e_c\ .\label{comm}
\end{equation}
Since coordinate bases commute, then $T^a$ measures to what extent
the basis $\{e_a(x)\}$ departs from a coordinate basis.

The theory makes contact with the metric tensor by declaring the
basis orthonormal:
\begin{equation}
g^{\mu\nu}\,e^a_\mu\,e^b_\nu\ =\ \eta^{ab}\, .
\end{equation}
Using the Eq.~(\ref{dual}), the metric is solved as
\begin{equation}
g_{\mu\nu}\ =\ \eta_{ab}\,e^a_\mu\,e^b_\nu\, ,\label{metric}
\end{equation}
so $e\, =\, \sqrt{-\det(g_{\mu\nu})}$. Although the metric
(\ref{metric}) is invariant under {\it local} Lorentz
transformations of the tetrad, the structure (\ref{comm}) and the
$f(T)$ action are preserved only under {\it global} Lorentz
transformations in the tangent space:
\begin{equation}
e_{a'}\ =\ \Lambda^a_{a'}\ e_a\ ,\ \ \ \ \ \ \ \ e^{a'}\ =\
\Lambda_a^{a'}\ e^a\ .
\end{equation}
Therefore, $f(T)$ theories are dynamical theories not just for the
metric (\ref{metric}) but for the entire tetrad. As will be
mention later, only in the special case of the so called
teleparallel equivalent of general relativity the theory acquires
invariance under local Lorentz transformations, so becoming a
dynamical theory just for the metric. In general, it could be said
that an $f(T)$ theory provides the spacetime not only with a
metric but with an absolute parallelization.

As a consequence of the lack of local Lorentz invariance, the
field of frames solving the dynamical equations can substantially
differ from a naive ``square root'' of the metric tensor, as tends
to be the guess when the metric is diagonal. In the case of
Friedmann-Robertson-Walker (FRW) universes, this naive ansatz only
works for spatially flat (K=0) cosmologies, but in
Ref.~\cite{Nos5} it was shown that the frames parallelizing
spatially curved FRW cosmologies are far from being the ``squared
root'' of the corresponding (diagonal) metric. For instance, in
the closed FRW model, whose topology is $R\times S^3$, the
preferred frame inherits the structure
$e^a=(dt,\overcirc{e}^{i})$, where the fields $\overcirc{e}^{i}$,
$i=1,2,3$ are responsible for the parallelization of the three
sphere $S^3$. In this way, the spatial part of $e^a$ turns out to
be highly non diagonal in the usual spherical coordinates. Other
less symmetric spacetimes, like the homogeneous and anisotropic
(open) Bianchi type I cosmologies, are however susceptible of a
more graceful treatment because their spatial topological
structure $R^3$ is expressed also as a product of parallelizable
submanifolds. In this case the tetrads with the form
$e^a=(dt,a(t)\,dx,b(t)\,dy,c(t)\,dz)$, being $a(t),b(t),c(t)$ the
(time dependent) scale factors, are certainly representative of
the parallelization process and they have been also used recently
\cite{fdet101}.

The parallelization involved in the dynamical fields $e^a$
represents a serious trouble for characterizing a solution on the
basis of the symmetry of the metric tensor. For instance,
hypersurfaces of constant $t$ in the Schwarzschild spacetime have
topology $R\times S^2$. This submanifold is three-dimensional, so
it is certainly parallelizable. However, the form of the parallel
fields is not trivial because $S^2$ is not parallelizable. So, in
spite of the simple structure of the metric tensor, the obtention
of the \emph{Schwarzschild frame}, i.e.~the frame field which
leads to the Schwarzschild solution in $f(T)$ theories, is a
difficult task that requires a more systematic approach not yet
developed. In particular, the recent study of spherically
symmetric static spacetimes in $f(T)$ theories
\cite{Wang,Deli,Boe,Meng,Houndjo,Houndjo2} did not get the correct
parallelization. Actually, vacuum solutions with spherical
symmetry are inconsistent with the frames considered in those
works, so the existence of Birkhoff's theorem in four-dimensional
$f(T)$ theories remains as an open question.

After reviewing the essentials of $f(T)$ theories in section
\ref{sec1}, we construct the \emph{Schwarzschild frame} in section
\ref{sec2}, which give rise to the Schwarzschild metric. This
special frame turns out to be a boost of the isotropic (diagonal)
frame, and represents the static, spherically symmetric vacuum
solution of every ultraviolet smooth deformation of general
relativity with $f(T)$ structure. In section \ref{sec3}, and in
the modest environment of the three-dimensional $f(T)$ gravity, we
show that circularly symmetric vacuum solutions are given by
BTZ-like black holes or ``Deser-de Sitter'' spacetimes (depending
on the sign of the effective cosmological constant that naturally
arises by virtue of the new scales present in the function
$f(T)$), provided the function $f(T)$ be of an infrared sort. In
the case of high energy $f(T)$ modifications, the general
circularly symmetric solution is given by the standard conical
geometry characterizing the ($\Lambda=0$) three-dimensional
Einstein theory. Both results together constitute the Birkhoff's
theorem for three-dimensional $f(T)$ gravities. Finally, we
display the conclusions in section \ref{sec4}.

\section{The $abc$ of $f(T)$ gravity}\label{sec1}

Very much alike the $f(R)$ gravitational schemes, where the scalar
curvature $R$ characterizing the Hilbert-Einstein Lagrangian is
replaced by an arbitrary function $f(R)$, the $f(T)$ gravities come
into the scene as a result of replacing the Weitzenb\"{o}ck scalar
$T$, which  is the cornerstone of the {\it teleparallel equivalent}
of GR \cite{Haya}, by an arbitrary function $f(T)$. The
Weitzenb\"{o}ck scalar $T$ is a quantity quadratic in the 2-forms
$T^a$, which are usually introduced as the torsion associated to the
Weitzenb\"{o}ck connection $\Gamma^\lambda_{\mu\nu}\,=\,
e^\lambda_a\,\partial_\mu e^a_\nu$. In fact,
\begin{equation}
T_{\ \ \mu \nu }^{\lambda }\equiv e_{a}^{\lambda }\ T_{\ \ \mu \nu
}^{a}\,=\,
\Gamma^\lambda_{\nu\mu}-\Gamma^\lambda_{\mu\nu}\,=\,e_{a}^{\lambda
}\,(\partial _{\nu }e_{\mu }^{a}-\partial _{\mu }e_{\nu }^{a})\ .
\label{torsion}
\end{equation}
It is not difficult to verify that Weitzenb\"{o}ck connection is
metric-compatible and has null curvature.

In Refs.~\cite{Haya,Hehl2} the more general Lagrangian quadratic in
$T^a$ has been displayed as a linear combination of the quadratic
scalars associated with the three irreducible parts of $T^a$ under
the Lorentz group $SO(1,3)$:
\begin{equation}\label{combinacioncuadratica}
L_{||}=\frac{1}{16 \pi
G}\,\Big[de^{a}\wedge\,^{\ast}(a_{1}\,^{(1)}T^a+a_{2}\,^{(2)}T^a+a_{3}\,^{(3)}T^a)\Big]\,
,
\end{equation}
where the tensorial, vectorial and axial-vectorial parts $^{(1)}T^a$
, $^{(2)}T^a$ and $^{(3)}T^a$ are
\begin{eqnarray}
^{(1)}T^a&=&T^a-\,^{(2)}T^a-\,^{(3)}T^a,\notag \\
^{(2)}T^a&=&\frac{1}{3}\,e^{a}\wedge(e_b\rfloor T^b),\notag\\
 ^{(3)}T^a&=&-\frac{1}{3}{}^*(\,e^{a}\wedge{}^*(T^{b}\wedge\,e_{b})). \label{partes}
\end{eqnarray}
In the last two equations $\rfloor$ is the interior product and
$\ast$ refers to the Hodge star operator. In general, these
Lagrangians are not invariant under local Lorentz transformation
of the tetrad, since $T^a$ acquires an additional term coming from
the derivatives of the transformation matrix $\Lambda
_{a}^{a^{\prime}}$:
\begin{equation}
T^{a}\longrightarrow T^{a^{\prime}} =\Lambda _{a}^{a^{\prime}} \,
T^{a}-e^{a}\wedge d\Lambda _{a}^{a^{\prime}}\, .  \label{change}
\end{equation}
Therefore these theories describe global frames rather than just a
metric tensor. Notice that the Weitzenb\"{o}ck covariant derivative
of a vector field $V^\nu$ is
\begin{equation}
\nabla_\mu V^\nu\, =\, \partial_\mu V^\nu+V^\lambda e_a^\nu \partial
e_\lambda^a\, =\, e^\nu_a\, \partial_\mu(V^\lambda e^a_\lambda).
\end{equation}
So, a global frame allows to call {\it constant} to those vector
fields keeping constant its projections on the frame. Of course,
such a notion of {\it constant} would not make sense in theories
admitting local Lorentz transformations of the frames. The selection
of global frames constitutes an intrinsic feature of these theories:
the spacetime is endowed with an ``absolute parallelism'' determined
by the grid of field lines of vectors $e_a$.

However a very peculiar choice of the coefficients in
Eq.~(\ref{combinacioncuadratica}),
\begin{equation}\label{valordecouplings}
a_{2}=-2a_{1},\,\,\,\,\,\,a_{3}=-a_{1}/2\, ,
\end{equation}
makes the Lagrangian invariant under local Lorentz transformations
in the tangent space. In such case the Lagrangian results (including
a cosmological constant term) \cite{Haya,Arcos}
\begin{equation}
L_{\mathbf{T}}[e^{a}]=\frac{1}{16\pi G}\;e\;(T-2\Lambda)\;,
\label{lagrangianT}
\end{equation}
where
\begin{equation}
T\ =\ S_{\ \mu \nu }^{\rho }\ T_{\rho }^{\ \mu \nu }\, ,
\label{Weitinvar}
\end{equation}
and
\begin{equation}
S_{\ \mu \nu }^{\rho }=\frac{1}{4}\,(T_{\ \mu \nu }^{\rho }-T_{\mu
\nu }^{\ \ \ \rho }+T_{\nu \mu }^{\ \ \ \rho })+\frac{1}{2}\ \delta
_{\mu }^{\rho }\ T_{\sigma \nu }^{\ \ \ \sigma }-\frac{1}{2}\ \delta
_{\nu }^{\rho }\ T_{\sigma \mu }^{\ \ \,\sigma }\;.  \label{tensor}
\end{equation}
The reason why this particular Lagrangian is invariant under local
Lorentz transformations is because the Weitzenb\"{o}ck scalar $T$
can be rewritten as the sum of a term depending just on the metric
(\ref{metric}) --which possesses local Lorentz invariance-- and a
(non-locally invariant) surface term:
\begin{equation}
T\, =\, -R + 2\; e^{-1}\;\partial _{\nu }(e\,T_{\sigma }^{\ \sigma
\nu }\,)\;.  \label{divergence}
\end{equation}
The term depending just on the metric is the Levi-Civita scalar
curvature $R$. This means that the theory (\ref{lagrangianT})
reveals itself as a theory completely equivalent to GR. This
teleparallel form of the Hilbert-Einstein Lagrangian is known as
the teleparallel equivalent of general relativity (TEGR)
\cite{Haya,Hehl2,Arcos,Maluf}. $f(T)$ theories of gravity are
deformations of the Lagrangian (\ref{lagrangianT}), so their
dynamics are governed by the action
\begin{equation}
\mathcal{I}=\frac{1}{16\pi G}\int d^{4}x\;e\;f(T)+\int d^{4}x\;e\;
L_{\mathcal{M}}\, ,  \label{action}
\end{equation}
where $L_{\mathcal{M}}$ refers to the scalar Lagrangian of matter.
Excepting for linear $f$´s, the action (\ref{action}) does not
possess local Lorentz invariance due to the behavior of the second
term in the expression (\ref{divergence}). This implies that
$f(T)$ theories involve more degrees of freedom than general
relativity \cite{Barrow,Miao}. This situation can be compared with
$f(R)$ theories in the metric formalism: although their dynamical
variable is just the metric tensor, it appears one additional
degree of freedom due to the fact that the dynamical equations are
fourth-order in $f(R)$ theories. Instead, $f(T)$ theories have
second order dynamical equations --since no second order
derivatives appear in the Lagrangian--, but their dynamical
variable --the tetrad or vierbein-- has more components than the
metric tensor.

By varying the action (\ref{action}) with respect to the vierbein
components $e_{\mu }^{a}(x)$ we obtain the dynamical equations
\begin{eqnarray}
&&e^{-1}\partial _{\mu }(e\,S_{a}^{\ \ \mu \nu })\ \,f_{T }+e_{a}^{\
\ \lambda }S_{\rho }^{\ \ \nu \mu }\ T_{\ \ \mu \lambda
}^{\rho }\,\ f_{T} + \notag \\
&&+S_{a}^{\ \ \mu \nu }\,\partial _{\mu }T\ \,f_{TT }+ \frac{1}{4}\
e_{a}^{\ \ \nu }\,f=4\pi G\,\ \mathbb{T}_{a}^{\ \ \nu }\, ,
\label{dynamics}
\end{eqnarray}
where $\mathbb{T}_{a}^{\ \ \nu }=e_{a}^{\ \, \mu}\, \mathbb{T}_{\mu
}^{\nu }$ refers to the matter energy-momentum tensor
$\mathbb{T}_{\mu \nu }$, and $f_{T}$, $f_{TT}$ are the first and
second derivatives of $f$.

Equations (\ref{dynamics}) hide a crucial and very important
property. Let us consider a vacuum solution $e_{a}^{\nu }(x)$ of
Einstein equations. This means that the tetrad $e_{a}^{\nu }(x)$
solves the equations (\ref{dynamics}) with $f(T)=T-2\Lambda$ and
$\mathbb{T}_{\mu}^{\ \ \nu}=0$, i.e., $e_{a}^{\nu }(x)$ satisfies
\begin{equation}
e^{-1}\partial _{\mu }(e\,S_{a}^{\ \ \mu \nu })+e_{a}^{\lambda
}S_{\rho }^{\ \ \nu \mu }\,T_{\ \ \mu \lambda }^{\rho }+
\frac{1}{4}\ e_{a}^{\nu }\,(T-2\Lambda)=0\, . \label{dynamicsGR}
\end{equation}
It is quite important to note that, in spite of the fact that
$e_{a}^{\nu }(x)$ is a \emph{vacuum} solution, it does not generally
lead to a null or constant Weitzenb\"{o}ck scalar $T$. This point
can be easily seen by contracting the Eq.~(\ref{dynamicsGR}) with
the inverse tetrad $e^{a}_{\nu}(x)$. In so doing, one obtains the
scalar equation
\begin{equation}
 e^{a}_{\nu}\ \partial_{\mu }(e\,S_{a}^{\ \ \mu \nu })=2\,e\,\Lambda\, , \label{ecescalar}
\end{equation}
which, in principle, does not compel the invariant $T$ to be null or
constant for vacuum solutions. However, by virtue of the relation
(\ref{divergence}) we can see that the scalar $T$ must reduce to a
total derivative in that case.

Since the Einstein theory or its teleparallel equivalent allows for
local Lorentz transformations of the tetrad, one can wonder whether
this GR solution can be rephrased as a frame
$\bar{e}^{a}(x)=\Lambda^a_b (x)\, e^{\,b}(x)$ making constant the
Weitzenb\"{o}ck scalar $T$, let us say
${T}[\bar{e}^a]=2\bar{\Lambda}$. Of course, the existence of such a
frame is independent of the chosen coordinates, and has not effect
at all in the spacetime metric because $g$ is given by the tetrad
through $g=\eta_{ab}\,e^a\otimes e^b$, which is Lorentz invariant.
According to Eq.~(\ref{divergence}), we are asking for a Lorentz
transformation $\Lambda^a_b (x)$ such that
\begin{equation}
\bar{\Lambda} = e^{-1}\,\partial _{\nu }(e\,{T}_{\sigma }^{\ \
\sigma \nu }[\bar{e}^a])\;, \label{transfdivergence}
\end{equation}
because ${R}[\bar{e}^a]=R[{e}^a]=0$ for vacuum solutions, and
$\bar{e}=e$. Let us suppose that we have found such a frame and
replace it in the deformed equations (\ref{dynamics}). If
$f_{T}(2\bar{\Lambda})\neq 0$, the result is
\begin{equation}
e^{-1}\partial _{\mu }(e\,S_{a}^{\ \ \mu \nu })+e_{a}^{\lambda }\
S_{\rho }^{\ \ \nu \mu }\ T_{\ \ \mu \lambda }^{\rho }+ \frac{1}{4}\
e_{a}^{\nu }\ \frac{f(2\bar{\Lambda})}{f_{T}(2\bar{\Lambda})}=0\,.
\label{dynamics2}
\end{equation}
Since $\bar{e}^{a}(x)$ is already a solution of the Einstein
equations (\ref{dynamicsGR}) with $T=2\bar{\Lambda}$, then it will
also solve the $f(T)$ vacuum equations provided that
\begin{equation}
2(\bar{\Lambda}-\Lambda)=\frac{f(2\bar{\Lambda})}{f_{T}(2\bar{\Lambda})}\,
.\label{coscos}
\end{equation}
This argument can be extended to non-vacuum solutions, by adding the
substitution $G\rightarrow G/f_{T}(2\bar{\Lambda})$. In this way, we
have the remarkable result that, if we can find in Einstein's theory
the (locally rotated) frame $\bar{e}^{\,a}(x)$ such that
${T}[\bar{e}^a]=2\bar{\Lambda}$, then this special frame will also
solve those $f(T)$ theories accomplishing the relation
(\ref{coscos}). For instance, a GR vacuum solution with
$\bar{\Lambda}=0=\Lambda$ also solves any ultraviolet deformation of
GR:
\begin{equation}  \label{condiciones}
f(T)=T+ \mathcal{O}(T^2),\,\,\,\text{i.e.},\,\,\, f(0)=0,\,\,\,
\,\,\,f^{\prime}(0)=1\, .
\end{equation}
In next section, we will take advantage of this remarkable property
to show that Schwarzschild geometry solves any $f(T)$ theory
fulfilling the condition (\ref{condiciones}).

\section{Spherical symmetry and the Schwarzschild frame}\label{sec2}
Despite some claims on the contrary \cite{Meng}, it is easy to
verify that the frame
\begin{eqnarray}
e^0&=&\left(1-\frac{2M}{r}\right)^{1/2}\ dt\, , \notag \\
e^1&=&\left(1-\frac{2M}{r}\right)^{-1/2}\ dr\, , \notag\\
e^2&=&r\ d\theta\, , \notag\\
e^3&=&r \ sin\,\theta\ d\varphi\, , \label{naive}
\end{eqnarray}
which certainly leads to the Schwarzschild interval
\begin{equation}
ds^2=\left(1-\frac{2M}{r}\right)\,
dt^2-\frac{dr^2}{1-\frac{2M}{r}}-\, r^2\, d\Omega^2\, ,
\label{metusual}
\end{equation}
is not a consistent solution of the $f(T)$ dynamical equations. In
fact the $r$-$\theta$ equation of motion (\ref{dynamics}) yields
\begin{equation}
f_{TT}\,(16 M^3-8M^2r-2Mr^2+r^3)=0\, ,
\end{equation}%
which clearly cannot be satisfied except in the case
$f(T)=T-2\Lambda$. Worse yet, if $f_{TT}\neq0$ the
Eq.~(\ref{metusual}) is not fulfilled not even if $M=0$; i.e., the
tetrad (\ref{naive}) results unsatisfactory even for the
description of Minkowski spacetime in arbitrary $f(T)$ theories.
Of course, the parallelization of Minkowski spacetime is generated
by the Cartesian basis $\{dx^a\}$. On the contrary, the frame
(\ref{naive}) generates circles, which are certainly not
autoparallel curves of flat spacetime.

Of course, this fact does not means that the Schwarzschild
solution is absent in $f(T)$ theories, but, in turn, that the
frame (\ref{naive}) is not correct for the description of such a
spacetime. Actually, in this section, we will show that the
Schwarzschild spacetime appears as solution of every $f(T)$ theory
satisfying (\ref{condiciones}), but in a somewhat tricky way.

To begin with the construction, let us take the spherically
symmetric Schwarzschild metric in isotropic coordinates, given by
\begin{equation}
ds^2=A(\rho)^2\, dt^2-B(\rho)^2\, \left(dx^2+dy^2+dz^2\right)\, ,
\label{metisotrop}
\end{equation}%
where the functions $A$ and $B$ depend on the radial coordinate
$\rho=\sqrt{x^2+y^2+z^2}$, and they are
\begin{equation}
A(\rho)=\frac{2\rho-M}{2\rho+M},\,\,\,\,\,\,\,B(\rho)=\Big(1+
\frac{M}{2\rho}\Big)^2\, . \label{funciones}
\end{equation}%
The isotropic chart covers just the exterior region of the
Schwarzschild spacetime, as results clear from the relation between
the coordinate $\rho$ and the radial coordinate $r$ of the
Schwarzschild gauge, which is
\begin{equation}
 \sqrt{r^2-2M r}+r-M=2\rho\, .\label{relrad}
\end{equation}%
We introduce now the \emph{asymptotic} frame, which is supposed to
be a good approximation to the actual (physical) frame just at
spatial infinity, when the spacetime has a Minkowskian structure.
This frame comes by taking the squared root of the metric
(\ref{metisotrop}), so it reads
\begin{eqnarray}
e^0&=&A(\rho)\,dt\, , \notag \\
e^1&=&B(\rho)\,dx\, , \notag\\
e^2&=&B(\rho)\,dy\, , \notag\\
e^3&=&B(\rho)\,dz\, . \label{euclideas}
\end{eqnarray}%
The frame (\ref{euclideas}), unlike (\ref{naive}), is particularly
useful as starting point because it captures the asymptotic
geometrical meaning of the parallelization process, reflected in
the fact that, at spatial infinity, we have the \emph{Minkowskian}
frame $e^a_{\mu}(\infty)=\delta^a_{\mu}$ which gives a null
torsion tensor. However, the \emph{asymptotic} frame is also
incapable to globally describe the Schwarzschild spacetime, as can
be checked by replacing it in the motion equations.

According to the results of the previous section, we have to look
for a Lorentz transformation in such a way that, after its action on
the frame (\ref{euclideas}), we be able to achieve a null
Weitzenb\"{o}ck scalar. Bearing in mind this, and the fact that the
Weitzenb\"{o}ck scalar coming from the tetrad (\ref{euclideas})
involves the functions $A(\rho)$ and $B(\rho)$ as well as its first
derivatives, we focus the quest in a radial boost depending solely
on the radial coordinate $\rho$. With the usual definitions
\begin{equation}
\gamma(\rho)=\Big(1-\beta^{2}(\rho)\Big)^{-\frac{1}{2}},\,\,\,\,\,
\beta(\rho)=v(\rho)/c\, , \label{defboost}
\end{equation}
we found that the most general boosted (radial) frame coming from
(\ref{euclideas}) yields
\begin{widetext}
\begin{eqnarray}
\bar{e}^0&=&A(\rho)\gamma(\rho)\,dt-\frac{B(\rho)}{\rho}
\sqrt{\gamma^{2}(\rho)-1}\,[x\,dx+y\,dy+z\,dz]\, , \notag \\
\bar{e}^1&=&-\frac{A(\rho)}{\rho}\,\sqrt{\gamma^{2}(\rho)-1}\,x
\,dt+B(\rho)\Big[(1+\frac{\gamma(\rho)-1}{\rho^2}x^2)\,dx+
\frac{\gamma(\rho)-1}{\rho^2}\,x
\,y\,dy+\frac{\gamma(\rho)-1}{\rho^2} \,x \,z\,dz\Big]\, ,\notag \\
\bar{e}^2&=&-\frac{A(\rho)}{\rho}\,\sqrt{\gamma^{2}(\rho)-1}\,y
\,dt+B(\rho)\Big[\frac{\gamma(\rho)-1}{\rho^2}\,x
\,y\,dx+(1+\frac{\gamma(\rho)-1}{\rho^2}y^2)\,dy+
\frac{\gamma(\rho)-1}{\rho^2}\,y
\,z\,dz\Big]\, ,\notag \\
\bar{e}^3&=&-\frac{A(\rho)}{\rho}\,\sqrt{\gamma^{2}(\rho)-1}\,z
\,dt+B(\rho)\Big[\frac{\gamma(\rho)-1}{\rho^2}\,x\,
z\,dx+\frac{\gamma(\rho)-1}{\rho^2}\,y\,
z\,dy+(1+\frac{\gamma(\rho)-1}{\rho^2}z^2)\,dz\Big]\, .\notag \\
\label{boosteadas}
\end{eqnarray}%
\end{widetext}
The expression for the Weitzenb\"{o}ck scalar corresponding to the
boosted frame (\ref{boosteadas}) is rather complicated and it is not
worth of be showed here. However, it can be easily checked that with
the form (\ref{funciones}) for the functions $A(\rho)$ and
$B(\rho)$, the Weitzenb\"{o}ck invariant becomes zero if and only if
the Lorentz factor $\gamma(\rho)$ is chosen according to
\begin{equation}
\gamma(\rho)=\frac{4\rho^2+M^2}{4\rho^2-M^2}\, .
\end{equation}
In fact, in this case the non-null components of $T_{\lambda\mu\nu}$
and $S_{\lambda\mu\nu}$ are
\begin{eqnarray}
T_{00\alpha}&=&S_{00\alpha}\ =\
-\frac{M}{\rho^3}\,\frac{A(\rho)}{B(\rho)}\,
x_\alpha\notag\\
T_{\alpha 0\beta}&=&2\, S_{\alpha
0\beta}\,-\,\frac{M}{\rho^2}\,\delta_{\alpha\beta}\ =\ \,\frac{2\,
M}{\rho^4}\, x_\alpha\, x_\beta\, -\,
\frac{M}{\rho^2}\,\delta_{\alpha\beta}\notag\\
T_{\alpha\alpha\beta}&=&\frac{M}{\rho^3}\,\frac{B(\rho)}{A(\rho)}\,
x_\beta\, \ \ \ \ \ \ \alpha\neq\beta\, ,\label{TS}
\end{eqnarray}
and those coming from the antisymmetric behavior ($x_\alpha$ stands
for $x, y, z$); then it can be easily verified that
$T=S^{\lambda\mu\nu}T_{\lambda\mu\nu}$ cancels out. It is evident
that the boosted frame (\ref{boosteadas}) asymptotically reduces to
the one given in Eq.~(\ref{euclideas}), which, by virtue of the
definition (\ref{defboost}), means that the boost velocity $v(\rho)$
asymptotically goes to zero. Moreover, as $\rho$ approaches the
black hole horizon located at $\rho_{h}=M/2$, the boost velocity
goes to the speed of light. It should be remarked that the
expression (\ref{boosteadas}) is not longer valid in the interior
region of the black hole, since the isotropic chart does not cover
that part of the spacetime. However, one can change coordinates in
order to cover the interior piece of the black hole without
affecting the fact that $T[\bar{e}]$ is null. For instance, we can
obtain the maximum analytical extension of the Schwarzschild
spacetime after performing a change to Kruskal coordinates by means
of the change law
\begin{equation}
\bar{e} ^a_{\mu^\prime}\ =\ \frac{\partial x^{\mu}}{\partial
x^{\mu^\prime}}\ \bar{e} ^a_{\mu}\ ,
\end{equation}
where $x^{\mu^\prime}$ refers to the Kruskal chart and $x^{\mu}$ to
the isotropic one. The $\bar{e} ^a_{\mu^\prime}$ would be, then, the
\emph{same} frame in a different coordinate system.

The existence of the \emph{Schwarzschild frame} (\ref{boosteadas})
automatically proves that the Schwarzschild spacetime is a
solution of every $f(T)$ theory satisfying the condition
(\ref{condiciones}). It is quite important to emphasize that,
despite that both frames (\ref{euclideas}) and (\ref{boosteadas})
lead to the same metric (\ref{metisotrop}), which is a trivial
fact in the context of GR, the only consistent frame (up to
\emph{global} Lorentz transformations) that reproduces the
Schwarzschild spacetime in the context of $f(T)$-like gravities is
the \emph{Schwarzschild frame} (\ref{boosteadas}).

\section{Circular symmetry in D=2+1 dimensions}\label{sec3}
In three-dimensional spacetime, the most general metric compatible
with circular symmetry can be cast in the form
\begin{equation}
ds^{2}=N^{2}(t,r)\
dt^{2}-\frac{Y^{2}(t,r)}{N^{2}(t,r)}dr^{2}-r^{2}\left( N^{\theta
}(t,r)\ dt+d\theta \right) ^{2}. \label{metrica}
\end{equation}%
A suitable dreibein field for the metric (\ref{metrica}) is given
by
\begin{eqnarray}\label{triada}
\bar{e}^{0} &=&N(t,r)\, dt\ ,  \notag \\ \bar{e}^{1}
&=&\frac{Y(t,r)}{N(t,r)}\, dr\ ,\notag
\\ \bar{e}^{2} &=&r\, N^{\theta }(t,r)\,dt + r\,d\theta \
.\label{driebein}
\end{eqnarray}
The Weitzenb\"{o}ck invariant for this frame reads
\begin{equation}
T=\frac{2\,(N^2)^\prime + r^3\,({N^\theta}^\prime)^2}{2\,r\, Y^2}\
,\label{inv3d}
\end{equation}
where the primes indicate differentiation with respect to the
radial coordinate. Note that no time derivatives are involved in
the expression (\ref{inv3d}), so all circularly symmetric
spacetimes are also stationary \footnote{Concerning this point,
see S. Deser and B. Tekin, Class. Quant. Grav. \textbf{20} (2003)
4877.}. Because of this reason, we can write the \emph{lapse} $N$
and the \emph{shift} $N^{\theta }$ just as functions of $r$. The
presence of the shift function in the off-diagonal term of this
``spherically'' symmetric line element is one of the typical
subtleties featuring three-dimensional gravity.

The Einstein (teleparallel equivalent) vacuum solution with the form
(\ref{driebein}) is
\begin{eqnarray}
&&N^{\theta}(r)=-\frac{J}{2\,r^{2}}\ ,\notag \\
&&N^{2}(r)=-M-\Lambda\,r^{2}+\frac{J^{2}}{4\,r^{2}}\ ,\notag \\
&&Y=1\ , \label{undefor}
\end{eqnarray}
$M$ and $J$ being two integration constants associated with the mass
and angular momentum respectively. When the cosmological constant
$\Lambda$ is negative, we have the spinning BTZ black hole
\cite{btz}. In the case of null and positive $\Lambda$, in turn, we
have the conical geometry first studied by Deser, Jackiw and 't
Hooft \cite{3D} and by Deser and Jackiw in Ref. \cite{3D2}. The key
point here is that the Weitzenb\"{o}ck invariant (\ref{inv3d}) is
constant for the particular GR solution (\ref{undefor}); Its value
is  $T=-2\Lambda$. This fact simplifies the analysis enormously,
because the triad (\ref{triada}) plays the role of the
``transformed'' frame mentioned in the construction made at the end
of section II, which means that the solution (\ref{undefor}) will be
also a solution of $f(T)$ theories satisfying Eq.~(\ref{coscos})
with ${\bar\Lambda}=-\Lambda$. We will show this point by solving
the motion equations (\ref{dynamics}) in an explicit way. For this,
we replace the ansatz (\ref{triada}) in (\ref{dynamics}), and find
three independent equations
\begin{eqnarray}
&&f-2\ f_{T}\,T\ =\ 0\ ,\notag \\
&&f_{T}\ =\ c_{1}\ Y\ ,\notag \\
&&f_{T}\ {N^\theta}^\prime\ =\ c_{2}\ \frac{Y}{r^3}\ ,
\label{ecdemov}
\end{eqnarray}%
where $c_{1,2}$ are two integration constants and $T$ is given in
Eq.~(\ref{inv3d}). If $f_{T}$ is non null, we divide the last two
equations to obtain (up to an additive constant which can be
eliminated by a coordinate change)
\begin{equation}\label{shift}
N^{\theta}=-\frac{J}{2\,r^{2}}\ ,\,\,\,\,\,J\equiv \frac{c_2}{c_1}\
,
\end{equation}
which gives the shift function of Eq.~(\ref{undefor}). It is
important to note that this result is independent of the specific
choice for the function $f(T)$.

In order to proceed, let us suppose first that $T=0$. In such case,
the first equation in (\ref{ecdemov}) is accomplished by any
ultraviolet deformation (\ref{condiciones}). Moreover, the equations
(\ref{inv3d}) and (\ref{shift}) completely determine the lapse
function:
\begin{equation}\label{lapse}
N^{2}(r)\ =\ M + \frac{J^{2}}{4\,r^{2}}\ ,
\end{equation}
which, essentially, is the second equation in (\ref{undefor}) with
$\Lambda=0$. Besides, the relation
\begin{equation}
f_{T}(0)\ =\ c_{1}\, Y \label{ecdemovdos}
\end{equation}
in Eq.~(\ref{ecdemov}) says that we can choose $Y=1$ with no loss of
generality. In this way, we see that the conical geometry
characterizing the elementary solution of Einstein's theory in
three-dimensional spacetime also emerges out as the solution of
\emph{any} ultraviolet $f(T)$ theory.

Although the last conclusion resembles the one obtained in 3+1
dimensions for the Schwarzschild frame, some differences
distinguishing GR in D=3+1 and D=2+1 should be here mentioned. While
$M$ in Schwarzschild solution characterizes the local geometry, $M$
and $J$ in the metric (\ref{metrica}) with $Y=1$,
\begin{eqnarray}
ds^{2}&=&\left[ d(M\ t+J\ \theta /(2M))\right] ^{2}- \frac{dr^{2}}{%
J^{2}/(4\,r^{2})+M^{2}} -\notag
\\ &&-\frac{r^{2}}{M^{2}}(J^{2}/(4r^{2})+M^{2})\ d\theta ^{2}\ ,
\label{metricareg}
\end{eqnarray}%
could be regarded as aspects of the global geometry. In fact, they
can be absorbed by performing the coordinate change
\begin{eqnarray}
\rho&=&M^{-2}(J^{2}/4+M^{2}r^{2})^{1/2}\, ,\notag
\\
\tau&=&M\ t+J\ \theta /(2M)\, ,\notag \\
\varphi&=&M\ \theta\, ,\label{cambios2}
\end{eqnarray}
given so the metric
\begin{equation}
ds^{2}=d\tau^{2}-d\rho^{2}-\rho^2\,d\varphi ^{2}, \label{metcono}
\end{equation}
where now we have $0\leq\varphi\leq 2 \pi M$ and the new time
coordinate $\tau$ is forced to a jump of $\Delta \tau=J/\pi M$ as a
consequence of the discontinuity of $\varphi$. This means that the
geometry is locally flat whatever $M$ and $J$ are. As a consequence,
the pair $M,J$ in Eqs.~(\ref{triada}, \ref{undefor}) with
$\Lambda=0$, characterizes different frames for the {\it same} local
geometry. Moreover, since $T$ is zero for all the $M,J$ family of
frames, the theory is unable to pick a frame and remove this
ambiguity. For instance, the frames (\ref{triada}) and
\begin{equation}
 e^0=d\tau,\,\,\,\,e^1=d\rho,\,\,\,\,e^2=\rho\, d\varphi,
\label{tresdiag}
\end{equation}
which are connected by the boost
\begin{eqnarray}\label{boost3d}
&&\bar{e}^0=\frac{e^0-J/(2M^2\rho)\,e^2}{\sqrt{1-J^2/(4M^4\,\rho^2)}}\ ,\notag\\
&&\bar{e}^1=e^1\ ,\notag \\
&&\bar{e}^2=\frac{e^2-J/(2M^2\rho)\,e^0}{\sqrt{1-J^2/(4M^4\,\rho^2)}}\,,
\end{eqnarray}
both lead to a null Weitzenb\"{o}ck scalar $T$, in contrast with
what happens in the four dimensional analogue (remember that, in
general, $T$ is not invariant under local Lorentz transformations of
$e^a$, see Eq. (\ref{divergence})). This freedom in the choice of
the triad could be removed in a more general theory, as the one
proposed in Ref.~\cite{Nos5} where $M$ and $J$ do characterize the
local geometry. From another point of view, the use of the metric
(\ref{metricareg}) in the (2+1)-Einstein equations implies the
energy-momentum tensor $T^{tt}\propto \mu\, \delta^2({\bf r})$,
$T^{ti}\propto J\, \varepsilon^{ij}\,
\partial_j \delta^2({\bf r})$, where $\mu=(1-M)/4$. This means
that a spinning massive particle is located at the origin $r=0$.
Although the spinning particle at the origin does not locally
curve the spacetime, it does confer global properties to the
manifold; while the presence of mass generates a wedge in
Minkowski space, the effect of spin is to give the spacetime a
kind of helical structure, for a complete rotation about the
source is accompanied by a shift in time \cite{3D}. This could be
regarded as a physical reason to have different triads for the
same local geometry.

Let us return to the problem posed in Eq.~(\ref{ecdemov}), for
considering the cases where $T$ is non-null. The first equation in
(\ref{ecdemov}) is an algebraic equation for $T$. At this point we
have to prescribe the function $f(T)$ in order to carry on with the
analysis. Let us consider the case of the infrared deformations
originally proposed to describe the late cosmic speed up
\cite{fdet1}:
\begin{equation}  \label{modelo}
f(T)=T+\frac{\alpha}{(-T)^{\beta}}\,.
\end{equation}
Thus, multiplying the first equation in (\ref{ecdemov}) by the non
null quantity $(-T)^{\beta}$ we obtain
\begin{equation}\label{tconst}
T=-\left[-\alpha(1+2\beta)\right]^{\frac{1}{1+\beta}}\, .
\end{equation}
Therefore the second equation in (\ref{ecdemov}) just says
\begin{equation}\label{funy}
\frac{1+\beta}{1+2\beta}=c_{1}\,Y,
\end{equation}
which means that we can make $Y=1$ by choosing the constant $c_{1}$.
As mentioned before, the function $N^{\theta}$ does not depend on
the form of $f(T)$, so its expression is again (\ref{shift}).
Finally, the Eq.~(\ref{inv3d}) says that the lapse function is
\begin{equation}\label{lapsof}
N^2(r)=-M-\frac{1}{2}\,\left[-\alpha(1+2\beta)\right]^{\frac{1}{1+\beta}}\,
r^2+\frac{J^2}{4\, r^2}\ ,
\end{equation}
Thus, the infrared deformations (\ref{modelo}) lead to an
asymptotically de Sitter or anti de Sitter spacetime, according to
the sign of the effective cosmological constant,
\begin{equation}
\Lambda\ =\ \frac{1}{2}\ \left[-\alpha(1+2\,
\beta)\right]^\frac{1}{1+\beta}\ .\label{lbda}
\end{equation}
We see that a small effective cosmological constant can be obtained
not only by considering $\alpha\approx0$, but also setting
$\beta=-1/2+\epsilon$, with $|\epsilon|<<1$. In this case, the
effective cosmological constant is positive:
$\Lambda\approx2\alpha^2\epsilon^2$. Although this choice for the
parameters $\alpha$ and $\beta$ is not expected to be meaningful for
the four-dimensional analogue, nonetheless it points out a very
interesting line of research.

Again we remark that the infrared deformed solution
(\ref{shift})-(\ref{lapsof}) can be straightforwardly obtained from
its GR partner (\ref{undefor}). Since the GR solution
(\ref{undefor}) has a constant Weitzenb\"{o}ck invariant, $T=-2\,
\Lambda$, then we can invoke the argument displayed at the end of
section \ref{sec1}. Using that $\bar{\Lambda}=-\Lambda$, then the
Eq.~(\ref{coscos}) for the function (\ref{modelo}) leads to the
Eq.~(\ref{lbda}) for the effective cosmological constant.

\section{Concluding Remarks}\label{sec4}

We have shown in section \ref{sec2} that the Schwarzschild
spacetime remains as solution of \emph{ultraviolet} deformations
of GR with $f(T)$ structure. The Schwarzschild frame
(\ref{boosteadas}), which emerges from a radial boost of the
isotropic frame of Eq.~(\ref{euclideas}), is the solution for any
$f(T)$ subjected to the conditions of Eq.~(\ref{condiciones}).
This result rules out the possibility of obtaining a deformed
(hopefully regular) black hole solution as a vacuum state of
$f(T)$ schemes. However, getting the Schwarzschild frame
constitutes the starting point in the search for potential
deformations of \emph{infrared} character, which are important
nowadays in connection with the accelerated expansion experienced
by the Universe, and also to explore new physics in the Solar
System. Additionally, it makes possible to formulate more
realistic models for stars, since the matching between the
exterior (Schwarzschild) frame and the interior one will bring
additional constraints in the physical quantities describing the
star, such as the energy density and the pressure of the fluid.
Finally, we should emphasize that the results developed in the
section \ref{sec2} of this article just applies in the static
case; the existence of monopolar radiation coming from the
additional degrees of freedom that certainly are present in $f(T)$
theories, and so, the very validity of Birkhoff's theorem in this
context, remains as an open problem to be worked in the future.

With the aim of clarifying some of these issues, in section
\ref{sec3} we have worked in the gentler setting of
three-dimensional gravity. We saw, by virtue of the time
independence of the Weitzenb\"{o}ck invariant (Eq.~(\ref{inv3d})),
that all circularly symmetric vacuum solutions of $f(T)$ theories
are also stationary. Moreover, the elementary circular symmetric
solution in vacuum for all high energy deformations is given by the
conical spacetime of Deser et al. This seems to be a very stringent
property of $f(T)$ theories, not shared by other ultraviolet GR
deformations with absolute parallelism structure (see the regular
spacetime obtained in the context of Born-Infeld gravity
\cite{Nos4}). We also found that the ``exterior'' region of this
spacetime can be described by a family of triad fields connected by
local Lorentz boosts in the $(t-\theta)$ plane, suggesting so, at
least in three spacetime dimensions, the existence of a local
subgroup of the full Lorentz group that officiates also as a
symmetry group of the theory. More work is needed in order to
determine this subgroup as well as its physical significance.

Concerning the infrared deformations in D=2+1, we saw that the
effect of having a $f(T)$ function other than $T-2\Lambda$,
translates into the presence of an effective cosmological constant
coming from the new scales present in the function $f(T)$. For
instance, in the extensively considered infrared model
$f(T)=T+\alpha(-T)^{-\beta}$, the solution acquires an effective
cosmological constant given in Eq.~(\ref{lbda}). If this
cosmological constant turns out to be negative, the solution
represent a BTZ-like black hole, while a positive $\bar{\Lambda}$
give rise to a ``Deser-de Sitter'' spacetime with shift and lapse
functions given in Eqs.~(\ref{shift}) and (\ref{lapsof})
respectively. We expect to achieve analogous results in four
dimensions, though the formal proof constitutes a matter of
current research.

\acknowledgements This research was supported by CONICET and
Universidad de Buenos Aires.

\end{document}